\documentclass[useAMS,usenatbib,epsfig, a4paper,amsmath,aps]{mnras}
 
\usepackage{float}
\usepackage{epsfig}
\usepackage{subfig}
\usepackage{verbatim}
\usepackage{nicefrac}
\usepackage{url}
\usepackage{mathrsfs}
\usepackage{amssymb}
\usepackage{amsmath}
\usepackage{pbox}

\title[Reconciling Type Ia Supernova Rates]
   {Reconciling Volumetric and Individual Galaxy Type Ia Supernova Rates}
\author[P.~Andersen et al.]
{P.~Andersen$^{1}$\thanks{email: \href{mailto:perandersen@dark-cosmology.dk}{\nolinkurl{perandersen@dark-cosmology.dk}}} and J. Hjorth$^{1}$\\ 
$^{1}$Dark Cosmology Centre, University of Copenhagen, Juliane Maries Vej 30, 2100 Copenhagen O, Denmark.
}

\date{\today}
\pubyear{2017}

\begin{document}
\maketitle

\begin{abstract}
Significant observational effort has been devoted to determining volumetric type Ia supernova rates at high redshifts, leading to clues about the nature of Ia supernova progenitors and constraints on the iron production in the universe. A complementary approach is to investigate type Ia supernova rates in individual, more nearby, galaxies. The popular A+B model for the specific supernova rate, while reliable for a wide range of galaxy properties, diverges for large specific star formation rates. Applying it outside its range of validity could lead to the prediction of excessive type Ia supernova rates. Moreover, the A+B model it is not directly derived from a delay time distribution. We here introduce a new model which is explicitly motivated by
a simple delay time distribution composed of a prompt and a delayed component. The model is in remarkably good agreement with current observational constraints. It yields a prompt fraction of 
$f_{p}= 0.11^{+0.10}_{-0.06}$ in agreement with results based on volumetric rates of type Ia supernovae at high redshift \citep{Rodney2014}. The model is tested against realistic star formation rates from the Illustris-1 simulation and is found to be self consistent in the asymptotic limits. An analytic function that encapsulates the features of the new model is shown to be in excellent agreement with the data. 
In terms of goodness of fit, the new model is strongly preferred over the A+B model. At $\log{({\rm sSFR})} \gtrsim -9$ there are no constraints from observations. Observations in this regime will further constrain the delay time distribution of type Ia supernovae at short delay times.
\end{abstract}


\section{Introduction}
\label{sec:intro}
Type Ia supernovae (SNe) play a central role in modern astronomy. As standardisable candles they have been used to determine the apparent accelerated expansion of the universe \citep{Riess1998,Perlmutter1999}. In the chemical evolution of the universe they act as the primary source of iron \citep{Tsujimoto1995}. Despite their prominence, the precise physics of the Ia SNe progenitor is unknown. This is often referred to as the progenitor problem. Different models for the progenitor predict different delay time distributions (DTD). The DTD parametrises the distribution of times between the formation of the progenitor and the subsequent SN explosion. In the limit where all star formation happens instantaneously, in a brief starburst, the DTD parametrisation will reflect the rate of SN explosions that follow the starburst. In practise, the star formation history is more complex, and the observed SN rate is a convolution of the DTD and the star formation rate. This complicates modelling of SN Ia rates which plays an important role in, e.g., cosmological astrophysics and chemical evolution modelling.

Different avenues can be taken when modelling SN rates. One is to measure volumetric SN rates as a function of redshift \citep{Graur2013, Maoz2014, Rodney2014} and fit an assumed DTD. This approach seems to suggest a DTD of the form DTD $\sim \tau^{-1}$, perhaps with a transition to becoming constant at early times \citep[e.g.,][]{Rodney2014}. This may be explained by a progenitor model with two distinct progenitors. These two progenitors are often referred to as a prompt component, with short delay times, and a delayed component, with longer delays between star formation and SN explosion. Another approach is to observe SN rates for individual galaxies as a function of star formation rate of the host galaxy, averaged over all observed galaxies. This approach does not formally define or assume a DTD, and instead assumes that the SN rate depends on physical parameters such as stellar mass and star formation rate of the host galaxy.
The consensus model for this approach is the A+B model \citep{Mannucci2005,Scannapieco2005}, which, in the simplest form, relates the
SN Ia rate linearly to the stellar mass and the star formation rate, i.e.,
\begin{equation}
\mathrm{SNR} = A\,\,M_* + B\,\,\mathrm{SFR},
\end{equation}
where SNR is the SN rate, $M_*$ is the stellar mass, and SFR is the star formation rate. $A$ and $B$ are then the two free parameters of the model, with units of SNe$\,$yr$^{-1}\,M_{\odot}^{-1}$ and SNe$\,$yr$^{-1}(M_{\odot}\,\mathrm{yr}^{-1})^{-1}$ respectively. Often the stellar mass is eliminated from the above equation by dividing through with $M_*$, which in turn yields 
\begin{equation}
\label{eq:aplusb}
\mathrm{sSNR} = A + B \,\, \mathrm{sSFR},
\end{equation}
where $\mathrm{sSNR} = \mathrm{SNR} / M_*$ is the specific supernova rate and $\mathrm{sSFR} = \mathrm{SFR} / M_*$ is the specific star formation rate.\\

In \cite{Scannapieco2005} the A+B model is motivated by an observed bimodal distribution of type Ia SNe luminosities; they argue that the brightest SNe are more prevalent in actively star forming galaxies while underluminous events occur primarily in galaxies with low star formation rates. \cite{Sullivan2006} additionally find that in star forming galaxies, type Ia SNe occur more often in the disk than in the bulge, which suggests that the Ia rate is somehow dependent on recent star formation. \cite{Mannucci2006} furthermore argue that their finding that only DTDs including both a prompt and a delayed component are consistent with observations strongly suggests that there exists two classes of Ia progenitors. \cite{Rodney2014} fit a DTD that includes a prompt and delayed component, and find that this model fits the $z<1$ volumetric rates well with a prompt fraction of $f_p=0.53^{\pm0.09 \, \pm 0.10}_{\mathrm{\,stat \,\,\,\,\,sys}}$. At redshifts of $z>1$ they find comparatively fewer Ia SNe, and a prompt fraction of $f_p=0.21^{\pm0.34 \, \pm 0.49}_{\mathrm{\,stat \,\,\,\,\,sys}}$, which they argue is an indication that prompt progenitors might be less numerous than the $z<1$ sample suggests. The picture emerging from these studies is that the mechanism responsible for Ia explosions has two channels, one prompt and one delayed, and that both contribute significantly to the observed supernova rate.\\

In this work we attempt to reconcile the approach using volumetric rates and the approach using individual rates. To do so we will assume a DTD functionally similar to the one adopted by \cite{Rodney2014}. From this DTD we derive a model for the supernova rate as a function of star formation rate, and compare it to the A+B model. Additionally, from the functional form of the new model we will develop a simple expression that will predict supernova rates in the high redshift and high specific star formation regime.\\

The paper is structured as follows. First, in section \ref{sec:models}, we derive a new model for the relation between sSNR and sSFR. In section \ref{sec:data} we discuss the currently available data and describe the data set used in this work.  In section \ref{sec:method} we discuss the complications in fitting the models to the data, and proceed to fit both our new model and an A+B model. The results, including goodness of fit, using reduced $\chi^2$ and Kolmogorov--Smirnov tests, are presented in section \ref{sec:results}.  In section \ref{sec:illustris} we validate that the model is self consistent in the limits of high and low sSFR using realistic star formation histories from the Illustris-1 simulation. Finally, in section \ref{sec:discussion}, we discuss the applicability of the models in the high sSFR/high redshift regime and provide recommendations for how the new model can be used in practice.

\section{Models}
For clarity, we denote all times in this work that relate to the DTD with $\tau$, all times related to SNR with $t$, while galaxy ages are denoted with an upper case $T$.
\label{sec:models}
\subsection{Piecewise model}
\subsubsection*{Delay Time Distribution}
There is strong evidence for a DTD that follows a power law of $\tau^{-1}$, see e.g. \cite{Maoz2011} or \cite{Maoz2014}, Fig.\ 8. In \cite{Rodney2014} a three-branched DTD is defined, as being zero up to 0.04 Gyr, then constant up to 0.5 Gyr, and finally declining as $\tau^{-1}$ for $\tau>0.5$ Gyr. This corresponds to some fraction of the SNe making up a prompt fraction, and some a delayed fraction. Formally, the DTD of \cite{Rodney2014} takes the form
\begin{equation}
\mathrm{DTD}(\tau) = 
\begin{cases}
0 & \text{for } \tau<\tau_0\\
k_1 & \text{for } \tau_0 \le \tau \leq \tau_1\\
k_2 t^{-1} & \text{for } \tau_1<\tau,\\
\end{cases}
\end{equation}
with $(\tau_0,\tau_1)$ = ($0.04$ Gyr, $0.5$ Gyr). For simplicity we neglect the zero part the DTD where $\tau<\tau_0$. This is equivalent to assuming that $\tau_0=0$. This has no strong impact as this DTD does not diverge as $\tau \rightarrow 0$. The DTD used throughout this work is therefore defined as
\begin{equation}
\label{eq:dtd}
\mathrm{DTD}(\tau) = 
\begin{cases}
k_1 & \text{for } \tau \le \tau_1\\
k_2 \tau^{-1} & \text{for } \tau_1<\tau.\\
\end{cases}
\end{equation}
In the following we will derive the SNR in three regimes. For all regimes we apply the definition
\begin{equation}
\label{eq:snr}
\mathrm{SNR}(t) = \int_{t_a}^{t_b} \mathrm{SFR}(t-\tau) \mathrm{DTD}(\tau)\mathrm{d}\tau,
\end{equation}
where the limits $t_a$ and $t_b$ are set such that $\mathrm{SNR}(t)$ is positive. As both $\mathrm{SFR}(t-\tau)$ and  $\mathrm{DTD}(\tau)$ are positive, this means that the limits must be set such that $t_b > t_a$.

\subsubsection{Short Duration Starburst at Early Times}
We consider a galaxy with age $T_a$ where the star formation takes place from $T_a$ to $T_a - \Delta T$ in a short starburst of duration $\Delta T$ where $\Delta T \ll T_a$. Since we observe at late times, only the $\tau^{-1}$ component of the DTD is relevant. This yields
\begin{align}
\begin{split}
\mathrm{SNR}(t) &= \;\;\;\; \int_{T_a-\Delta T}^{T_a} \mathrm{SFR}(t-\tau) k_2 \tau^{-1} \mathrm{d} \tau\\
 &= -\int_{T_a}^{T_a-\Delta T} \mathrm{SFR}(t-\tau) k_2 \tau^{-1} \mathrm{d} \tau\\
&= -k_2 \mathrm{SFR}\ln{\frac{T_a-\Delta T}{T_a}} \approx k_2 \mathrm{SFR} \frac{\Delta T}{T_a}\\
&=\frac{k_2}{T_a} \, M_*,
\end{split}
\end{align}
where in the third line we Taylor expanded to first order and in the fourth line used that the product of the star formation rate and a time interval is the stellar mass produced. Additionally we assumed that the SFR is constant over the integrated time period, which is a valid assumption as $\Delta T \ll T_a$.

\subsubsection{Recent Starburst}
For a recent starburst at late times where $\tau<\tau_1$ the constant term of the DTD dominates, yielding
\begin{equation}
\label{eq:recentstarburstintegral}
\mathrm{SNR}(t) = \int_{0}^t \mathrm{SFR}(t-\tau) k_1 \mathrm{d} \tau = \mathrm{SFR} \, k_1 \, t,
\end{equation}
when assuming that the SFR is constant over the integrated time, which for a recent starburst is a good approximation. Dividing equation \ref{eq:recentstarburstintegral} through by the mass $M_*$ it yields
\begin{align}
\begin{split}
\label{eq:recentderive}
\mathrm{sSNR} &= \mathrm{sSFR} \, \, k_1 \, \, t.\\
\end{split}
\end{align}
In the recent starburst limit $sSFR \approx t^{-1}$. This follows from the relation between the mass $M_*$ and $SFR$,
\begin{equation}
M_* = \int_{0}^t \mathrm{SFR} \, \mathrm{d} t = \mathrm{SFR} \, t,
\end{equation}
and from the definition of the sSFR ,
\begin{equation}
\begin{split}
\label{eq:ssfrt1}
\mathrm{sSFR} &= \frac{\mathrm{SFR}}{M_*}\\
&= t^{-1}.
\end{split}
\end{equation}
Inserting the result of equation \ref{eq:ssfrt1} into equation \ref{eq:recentderive} yields
\begin{align}
\begin{split}
\mathrm{sSNR} &= k_1
\end{split}
\end{align}
for $\tau<\tau_1$, i.e., the specific supernova rate tends to a constant at high sSFR. $k_1$ can in principle be measured directly by counting the number of SNe in high sSFR galaxies. This also gives a direct measurement of the prompt fraction, defined as
\begin{align}
\begin{split}
f_p &= \frac{\int_{0}^{t_1} k_1 \mathrm{d}t}{\int_{0}^{t_1} k_1 \mathrm{d}t + \int_{t_1}^{t_2} k_2 t^{-1} \mathrm{d}t} \\
&= \frac{k_1 t_1}{k_1 t_1 + k_2\ln{(\frac{t_2}{t_1})}}
\end{split}
\label{eq:promptfraction}
\end{align}
where we follow \cite{Rodney2014} in denoting the prompt fraction $f_p$.

\subsubsection{Intermediate Times}
Above we derived the SNR in the early and late asymptotic limits. At intermediate times both the prompt and delayed component contribute. If we, for the sake of argument, assume that the SFR can be treated as a constant, then
\begin{align}
\begin{split}
\mathrm{SNR}(t) &= \int_{0}^{t_2} \mathrm{SFR}(t-\tau) \mathrm{DTD}(\tau)\mathrm{d}\tau\\
&= \int_{0}^{t_1} \mathrm{SFR}(t-\tau) k_1 \mathrm{d}\tau + \int_{t_1}^{t_2} \mathrm{SFR}(t-\tau) k_2 \tau^{-1} \mathrm{d}\tau\\
&= \mathrm{SFR} \, \, (k_1 t_1 + k_2 \ln{\frac{t_2}{t_1}}).
\end{split}
\end{align}
Dividing through with the mass $M_*$ gives
\begin{align}
\label{eq:intermediatetimes}
\mathrm{sSNR}(t) &= \mathrm{sSFR} \, \, \left(k_1 t_1 + k_2 \ln{\frac{t_2}{t_1}} \right).
\end{align}
The result of equation \ref{eq:intermediatetimes} is simply the product of the sSFR and a constant. The assumption of constant SFR is poor at intermediate times, so in order to generalise the behaviour we assume a power law in sSFR of the form
\begin{align}
\label{eq:ssfrpowerlaw}
\mathrm{sSNR(sSFR)} &= \alpha \,\, \left(\frac{\mathrm{sSFR}}{1 \, \mathrm{yr}^{-1}}\right)^\beta,
\end{align}
where in order to have the correct units a normalisation constant of size 1$\,$yr$^{-1}$ is introduced and $\alpha$ carries units of SNe$\,$yr$^{-1}\,M_{\odot}^{-1}$.
\subsubsection{Putting it all Together}
In the previous three sections we have suggested that the SNR is constant in the asymptotic limits and a power law at intermediate times. To parametrise this requires four free parameters, when imposing the requirement of continuity and fitting the for the transition points between the asymptotic limits and intermediate times. These four parameters are two parameters determining the constant SNR in the asymptotic limits and two parameters to fix the transition between the limits and intermediate times. To simplify the equations we utilise an equivalent but simpler parametrisation that has the transition points and then $\beta$ and $\alpha$ from equation \ref{eq:ssfrpowerlaw} as free parameters. Combining this in a continuous piecewise function yields
\begin{equation}
\label{eq:piecewise}
\mathrm{sSNR(sSFR)} = \alpha
\begin{cases}
S_1 & \text{for } \mathrm{sSFR}<\mathrm{sSFR}_2\\
 \left(\frac{\mathrm{sSFR}}{1 \, \mathrm{yr}^{-1}}\right)^\beta & \text{for } \mathrm{sSFR}_2<\mathrm{sSFR}<\mathrm{sSFR}_1\\
S_2 & \text{for } \mathrm{sSFR}>\mathrm{sSFR}_1,\\
\end{cases}
\end{equation}
where 
\begin{align}
&S_1 =  \left(\frac{\mathrm{sSFR}_1}{1 \, \mathrm{yr}^{-1}}\right)^\beta, \\
&S_2 =  \left(\frac{\mathrm{sSFR}_2}{1 \, \mathrm{yr}^{-1}}\right)^\beta,
\end{align}
and the four parameters to be fit are sSFR$_1$, sSFR$_2$, $\beta$, and $\alpha$.\\

\subsection{Smooth logarithm parametrisation}
The model of equation \ref{eq:piecewise} is not convenient for purposes that require a continuous differentiable model. Therefore we introduce a parametrisation that contains the same features, but is a smooth function for all values of sSFR. It is defined as
\begin{equation}
\mathrm{sSNR(sSFR)} = a + \frac{a}{k} \log{\left(\frac{\mathrm{sSFR}}{\mathrm{sSFR}_0} + b \right)},
\end{equation}
where the four free parameters to be fit are $a$, $k$, sSFR$_0$, and $b$.

\section{Data}
\label{sec:data}
The available type Ia supernova versus sSFR data include that of \cite{Mannucci2005}, \cite{Sullivan2006}, \cite{Smith2012}, \cite{Gao2013}, \cite{Graur2015a}, and \cite{Botticella2017}.
\cite{Sullivan2006} use type Ia observations from the Supernova Legacy Survey (SNLS) in the redshift range 0.2 $< z<$ 0.75. In \cite{Smith2012} type Ia SN observations are taken from the Sloan Digital Sky Survey (SDSS-II) SN Survey \citep{Frieman2008} which observed at redshifts of $z<0.4$. The data of \cite{Sullivan2006} and \cite{Smith2012} are analysed in a fashion that makes it straightforward to combine the two data sets. This includes assuming the same cosmology, both using AB magnitudes, binning and treating their data in a similar fashion. Other data sets exist, including those of \cite{Mannucci2005}. \cite{Mannucci2005} uses the SN observations of \cite{Cappellaro1999}. \cite{Gao2013} use spectroscopic data from the SDSS-II sample. \cite{Graur2015a} use a SN sample from SDSS Data Release 9. The most recent data is from the SUDARE survey \citep{Botticella2017}.\\

To work with data based on fully consistent assumptions and treatment of the data we choose to use the combined data set of \cite{Sullivan2006} and \cite{Smith2012} in the analysis of this work. The datapoint contaminated by systematically uncertain "ridge-line" galaxies of \cite{Smith2012} is on the basis of this contamination excluded from the analysis; see Fig. 5 of \cite{Smith2012} for details.\\

\section{Method}
\label{sec:method}
For all models we fit for maximum likelihood parameters by minimising the $\chi^2$ function given by
\begin{equation}
\chi^2(\mathrm{sSFR};P) = \sum_i \left( \frac{\mathrm{sSNR_{model}}(\mathrm{sSFR};P) - \mathrm{sSNR}_{\mathrm{data},i}}{\sigma_i} \right)^2
\end{equation}
where $\mathrm{sSNR_{model}}(\mathrm{sSFR};P)$ is the sSNR predicted by the model in question given parameters $P$ and $\mathrm{sSNR}_{\mathrm{data},i}$ is the $i$'th measurement with associated uncertainty $\sigma_i$. In practise what is done is we maximise the log-likelihood function which is related to $\chi^2$ via
\begin{equation}
\label{eq:loglikehood}
\log{\mathcal{L}(\mathrm{sSFR};P)} = \frac{-\chi^2(\mathrm{sSFR};P)}{2}.
\end{equation}
To fit the A+B model, the {\sc emcee} \citep{Foreman-Mackey2013} Markov chain Monte Carlo (MCMC) implementation was used to determine the maximum likelihood parameters utilising the log-likelihood function of Eq.~\ref{eq:loglikehood}. For the new models developed in this work the {\sc emcee} MCMC approach is unfortunately not viable. The reason is that the likelihood space contains a large number of local extrema. This causes the peaks in the one dimensional marginalised likelihoods to be a combination of different extrema of the likelihood space for different parameters, resulting in parameters that are a poor fit to the model. Therefore another approach is used for these models where the log-likelihood is calculated for values in a discrete grid in the parameter space of the model in question. From the grid of likelihoods the global maximum likelihood parameters are then extracted. Since the likelihood space is highly degenerate it is difficult to associate simple uncertainties to the maximum likelihood parameters without imposing strict priors, which would result in prior driven uncertainties. We apply the Fisher information matrix method (see \cite{Albrecht2006} or \cite{King2014} for an introduction to Fisher matrix methodology) to estimate uncertainties on the maximum likelihood parameters. The Fisher matrix method is a second order approximation of the likelihood and assumes Gaussian measurement uncertainties. The Fisher matrix method will yield the best theoretically possible statistical uncertainties, and neglects contributions from non-Gaussianity and systematics. More details on how the models introduced in this work are fit and links to the fitting and analysis code can be found in appendix \ref{app:fittingdetails}.\\

\begin{table*} 
 \centering 
 \begin{tabular}{| c | l | c | c | c | c |} 
 \hline Model & Maximum Likelihood Parameters & Reduced $\chi^2$  & BIC & AIC & KS\\
\hline A+B &  $A=(4.66 \pm 0.56) \cdot 10^{-14}$, $B=(4.88^{+0.54}_{-0.52}) \cdot 10^{-4}$& 3.5 & 36.3 & 35.5 & 0.21 \\ \\
Piecewise & \vtop{\hbox{\strut $\beta=0.586 \pm 0.084$, $\alpha=(1.19 \pm 2.20) \cdot 10^{-7}$,} \hbox{\strut sSFR$_2=(1.01 \pm 0.55) \cdot 10^{-11}$, sSFR$_1=(1.04 \pm 0.41) \cdot 10^{-9}$}}  & 1.2 & 18.0 & 16.4 & 0.048 \\ \\
Smooth logarithm &  \vtop{\hbox{\strut$a=(1.12 \pm 2.34)\cdot10^{-13}$, $k=0.49 \pm 1.53$,} \hbox{\strut  sSFR$_0=(1.665 \pm 0.002) \cdot 10^{-10}$, $b=0.73 \pm 0.99$}}   & 1.3 & 18.4 & 16.8 & 0.036  \\ \hline

 \end{tabular}  
\caption{Tabulated maximum likelihood parameter values and test values for the A+B model and the three models developed in this work. The two models from this work all have tests values that are comparable. This is unsurprising, as the models by construction contain the same functional features. The A+B model however consistently produce worse test values than the other models. The BIC and AIC tests are of special interest here as they penalise models for including additional fitting parameters, but despite this penalty the new models still produce significantly better test results than the A+B model. The units of $A$ and $B$ are SNe$\,$yr$^{-1}\,M_{\odot}^{-1}$ and SNe$\,$yr$^{-1}(M_{\odot}\,\mathrm{yr}^{-1})^{-1}$ respectively. For the piecewise model sSFR$_1$ and sSFR$_2$ carry units of yr$^{-1}$, $\alpha$ has units of SNe$\,$yr$^{-1}\,M_{\odot}^{-1}$ while $\beta$ is unitless. For the smooth logarithm $a$ and sSFR$_0$ have units of yr$^{-1}$ while $k$ and $b$ are unitless.}  
\label{tab:modelfits} 
\end{table*}

To test the quality of the model fits we calculate the reduced $\chi^2$, Bayesian information criterion (BIC), Akaike information criterion (AIC), and perform the Kolmogorov-Smirnov (KS) test for each model fit. The reduced $\chi^2$ is defined as\\
\begin{equation}
\chi^2_{\mathrm{reduced}} = \frac{\chi^2}{\nu}
\end{equation}
where $\nu$ represents the degrees of freedom, given by $\nu = N - n_p$ where $N$ is the number of data points and $n_p$ is the number of free parameters. The BIC is defined as
\begin{equation}
BIC = \chi^2 + n_p \,\cdot \, \log{N}
\end{equation}
and the AIC as
\begin{equation}
AIC = \chi^2 + 2\,n_p.
\end{equation}
 \begin{figure}
  \centering\includegraphics[width=\linewidth]{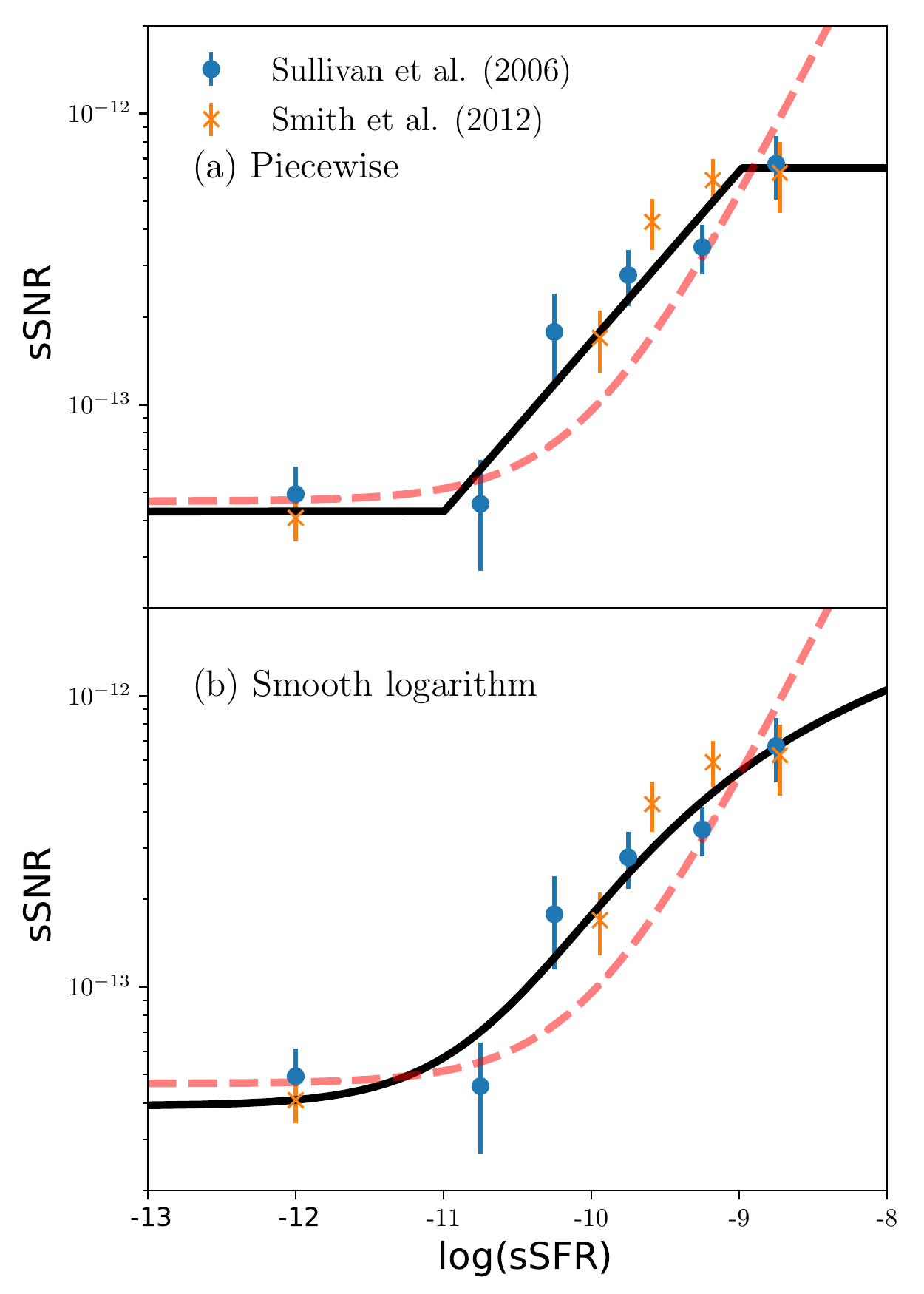}
  \caption{The piecewise and smooth logarithm maximum likelihood parameter fits plotted against the \protect\cite{Sullivan2006} and \protect\cite{Smith2012} data as well as the maximum likelihood AB model plotted as the dashed red line. The used parameters are shown in table \ref{tab:modelfits}. The plots show that the two models defined in this work are functionally similar, and that the A+B model increasingly deviates from these models at larger values of sSFR. As there is no data to constrain models at $\log{(\mathrm{sSFR})}>-9$ the sSNR in this regime is associated with large uncertainty and is strongly model dependent. In terms of goodness of fit (results shown in table \ref{tab:modelfits}) both the piecewise and smooth logarithm models outperform the A+B model.}
  \label{fig:newmodels}
\end{figure}

\section{Results}
\label{sec:results}
\begin{figure}
  \centering\includegraphics[width=\linewidth]{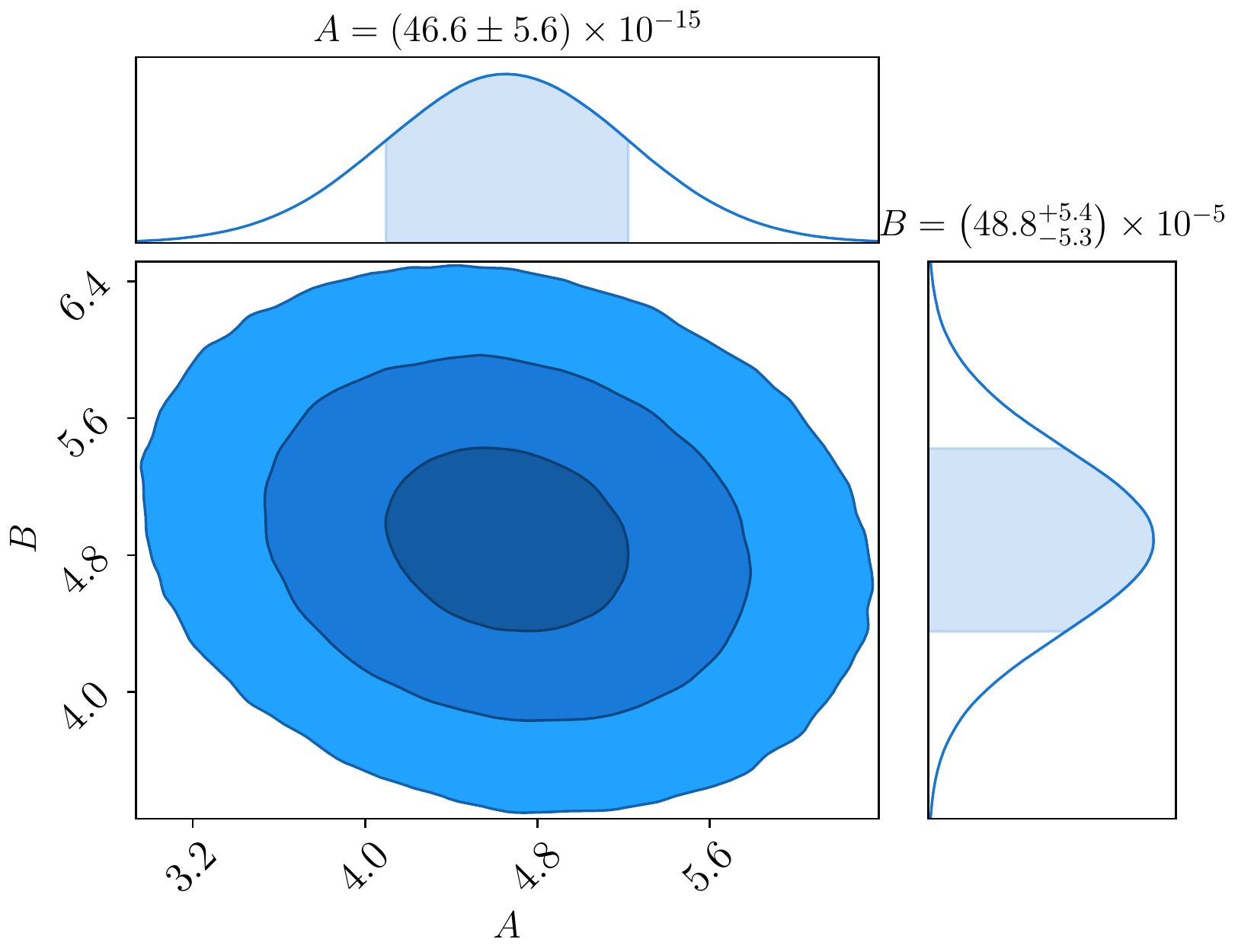}
  \caption{The corner plot from fitting the A+B model to the combined data set of \protect\cite{Sullivan2006} and \protect\cite{Smith2012} using the {\sc Emcee} MCMC tool. It is apparent from the contour plot, showing one, two, and three sigma uncertainties, that there is some mild covariance between the fitted parameters. The units of $A$ and $B$ are SNe$\,$yr$^{-1}\,M_{\odot}^{-1}$ and SNe$\,$yr$^{-1}(M_{\odot}\,\mathrm{yr}^{-1})^{-1}$ respectively. This figure was created using the {\sc ChainConsumer} package of \protect\cite{Hinton2016}.}
  \label{fig:simple}
\end{figure}

We apply the method of section \ref{sec:method} to produce the fits shown in figure \ref{fig:newmodels}. The details of the fits and test values are listed in table \ref{tab:modelfits}.\\

The derived maximum likelihood parameters of $A=(4.66 \pm 0.56) \cdot 10^{-14}$ and $B=(4.88^{+0.54}_{-0.52}) \cdot 10^{-4}$ for the A+B model are consistent with those in the literature, see e.g. figure 1 of \cite{Gao2013}. Further fit details are available in figure \ref{fig:simple}. Table \ref{tab:modelfits} shows that the A+B model provides the worst fit of the ones tested in this work for all test cases; reduced $\chi^2$, BIC, AIC, and KS test. Especially the BIC and AIC are of interest here, as they contain a penalty for including additional free parameters; in this way they test whether including the additional parameters is justified. Both models developed in this work yield test values that are similar within the random noise that is expected when deriving reduced $\chi^2$ values. This noise is simply a product of the uncertainty in each measurement, which in turn causes the reduced $\chi^2$ value itself to be uncertain. It is unsurprising that the two models developed in this work yield so similar test results, as they by construction contain the same overall functional shape. From figure \ref{fig:newmodels} this constructed similarity is apparent.\\

Our findings are in agreement with \cite{Smith2012} who find that the assumed linear relation of the A+B model provides a poor fit to data. Rather they find that a relation of $\mathrm{SNR} \propto M_*^{0.68}$ is preferred. \cite{Gao2013} furthermore find that the A+B model provides a poor fit to their data and show that there exists tension between A+B model parameter estimates in the literature, although the tension is weak as the uncertainties are relatively large (see Figure 1 of \cite{Gao2013}). Both of these studies indicate that there is some behaviour that the A+B model does not capture, and that additional fitting parameters are needed to adequately model the Ia supernova rates.\\

\section{Comparison with Simulations}
\label{sec:illustris}
 \begin{figure}
  \centering\includegraphics[trim={0cm 0 0cm 0},width=\linewidth]{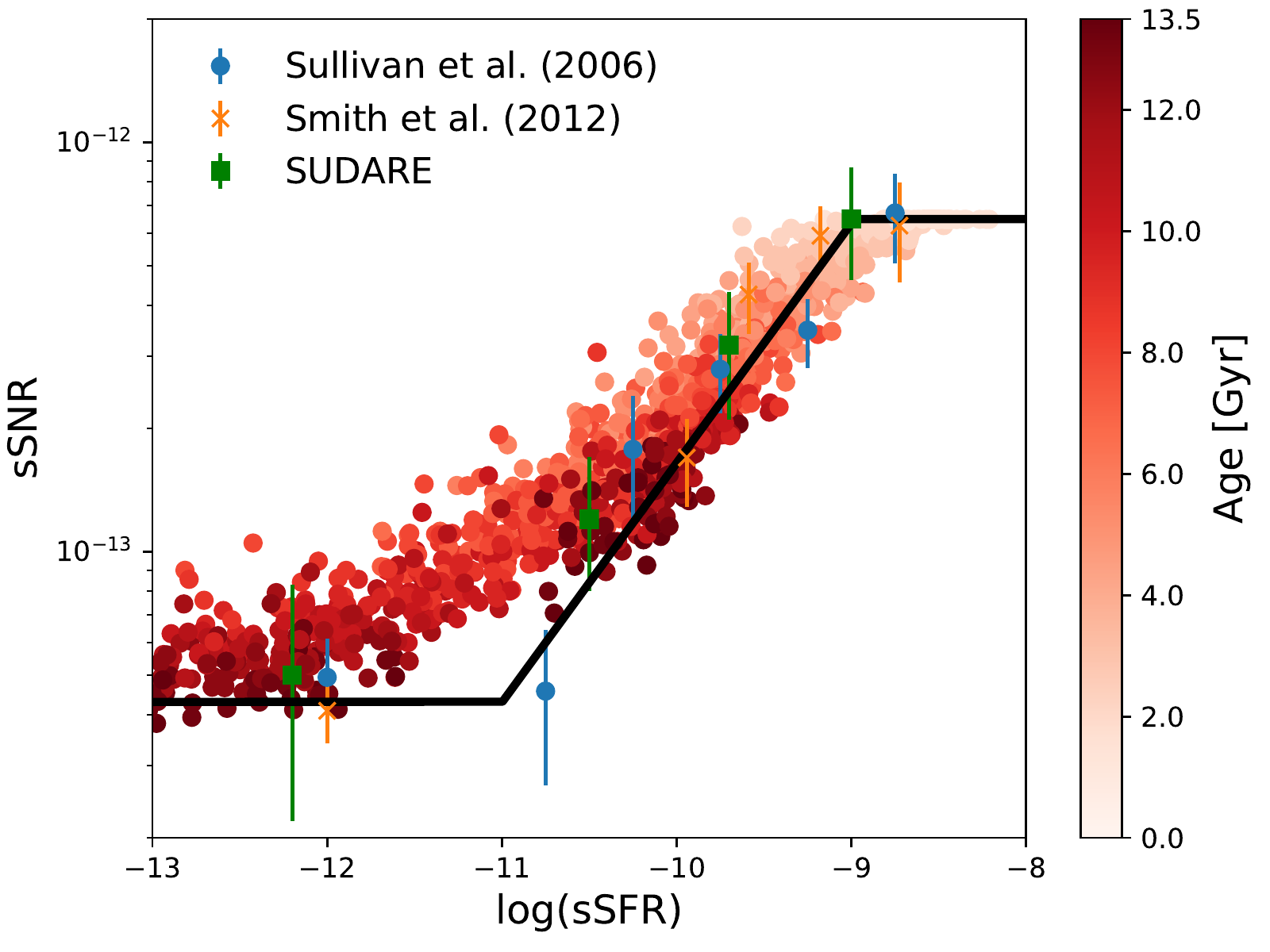}
  \caption{The red shaded dots show sSFR versus sSNR for star formation histories of simulated Illustris galaxies at discrete times in the range 1 Gyr to 13.7 Gyr. The colourbar shows the relation between colour and age, the redder the older the galaxy. The specific supernova rate is calculated for each galaxy using equation \ref{eq:snr} where the DTD of equation \ref{eq:dtd} is assumed with the best fitting parameters for the piecewise model from table \ref{tab:modelfits}. The observations of \protect\cite{Sullivan2006}, \protect\cite{Smith2012}, and SUDARE \protect\citep{Botticella2017} as well as the best fitting piecewise model to the observation of \protect\cite{Sullivan2006} and \protect\cite{Smith2012} is overplotted for comparison.}
  \label{fig:illustris}
\end{figure}

Assuming constant sSFR in section \ref{sec:models} will not hold true for all regimes of the piecewise model. In this section we test what the transition between the asymptotic limits looks like for realistic star formation histories. The applied approach is to determine the sSNR and sSFR  for a number of galaxies with more realistic star formation rates using equation \ref{eq:snr} and compare them with the predictions of the piecewise model. A source of realistic star formation histories can be found in the Illustris simulation \citep{Vogelsberger2014,Vogelsberger2014a,Genel2014}. Illustris is a hydrodynamical simulation going beyond simulating only the gravity of dark matter particles and also includes baryonic physics. Modelling gas and galaxies using computational fluid dynamics enables the Illustris simulation to predict the properties of the galaxies in the (106.5 Mpc)$^3$ large simulated volume. From 29,276 Illustris-1 galaxies with $M_*>10^9 M_\odot$ a subsample of 100 galaxies is chosen such that galaxies from all mass ranges are represented. This data collection is based on the analysis tools from \cite{Sparre2015} and \cite{Diemer2017}. For each galaxy the SFR as a function of time is extracted, and convolved in equation \ref{eq:snr} with the DTD of equation \ref{eq:dtd} to produce supernova rates at discrete times ranging from 1 Gyr to 13.7 Gyr. Deriving the mass of the galaxy at each of these discrete times using
\begin{equation}
\label{eq:stellarmass}
M_* = \int_0^t {\rm SFR}(t') \mathrm{d}t'
\end{equation}
allows us to determine the specific star formation and specific supernova rates. The results are shown as the red shaded dots in figure \ref{fig:illustris}. The shading of the dot indicates age, with darker shades meaning older galaxies. It is apparent that in the limits of very high or very low specific star formation rate the Illustris galaxies tends to the same specific supernova rates as the piecewise model. The exact values of specific supernova rate that the sample tends towards is a function of the assumed DTD parameters. If higher or lower $k_1$ and $k_2$ parameters were given as input in equation \ref{eq:dtd} this would be reflected in the limits the galaxies in figure \ref{fig:illustris} tend towards. As the piecewise function is fit to the observed data it captures the features therein, rather than the features of the Illustris galaxies. When the observations deviate from the Illustris results the piecewise model that best fits these observations will do the same. The reason for the slight discrepancy between Illustris and observations could be that a significant fraction of the stellar mass of galaxies is accreted rather than formed in situ. This additional mass from accretion is not taken into account when using equation \ref{eq:stellarmass} to determine the stellar mass. Figure 3 of \cite{Rodriguez-Gomez2016} shows that the accreted fraction is in the range of 0\% for galaxies of mass $10^9 M_\odot$ to 80\% for galaxies of mass $10^{12} M_\odot$. Another effect which can influence the mass of the galaxies is mass loss through massive stars dying and leaving remnant stars with masses smaller than the original ones. Figure 110 of \cite{Leitherer1999}, figure 3 of \cite{Courteau2014}, or equation 14 of \cite{Behroozi2013} show that on timescales of gigayears at most ${\sim}50$\% of the mass is lost due to this effect. In summary, the effects of mass loss or accretion will  typically yield a factor two increase or decrease in mass. Considering the effects discussed above and the statistical uncertainties on the observed data, the agreement between observations and the Illustris results is remarkably good.
\section{Discussion}
\label{sec:discussion}
The DTD used in this work is functionally identical to the one of \cite{Rodney2014}. Therefore it is possible to compare the value of the prompt fraction, $f_p$, from our fit to that of \cite{Rodney2014}. In \cite{Rodney2014} the transition from a constant to $\tau^{-1}$ DTD is fixed at 0.5 Gyr. Following the prescription of equation \ref{eq:promptfraction}, making it explicit that we fix the transition to be 0.5 Gyr with a subscript of 0.5, we arrive at a prompt fraction of $f_{p,0.5}= 0.11^{+0.10}_{-0.06}$ which is in slight tension with the result of $f_p=0.53^{\pm0.09 \, \pm 0.10}_{\mathrm{\,stat \,\,\,\,\,sys}}$ where \cite{Rodney2014} utilise their full data set, but is in good agreement with the result of $f_p=0.21^{\pm0.34 \, \pm 0.49}_{\mathrm{\,stat \,\,\,\,\,sys}}$ where only the CANDELS+CLASH sample of Ia supernovae was used. If we do not fix the transition to happen at 0.5 Gyr, and instead use the maximum likelihood fit value of $t_1= 964_{-273}^{+631}$ Myr we find a prompt fraction of $f_{p}=0.15^{+0.11}_{-0.09}$. This transition time is significantly larger than the fixed value of 0.5 Gyr of \cite{Rodney2014} but considering the uncertainties there is only slight tension.\\

From figure \ref{fig:newmodels} it is apparent that the difference in predicted Ia supernova rate between the A+B model and the models developed in this work increase with sSFR. The fit suggests that the biggest shortcoming of the A+B model is that it diverges towards increasingly larger values of supernova rate with sSFR, where the models that instead plateau at these large sSFR values in our analysis perform better. This suggests that the A+B model and the assumed DTD of \cite{Rodney2014} are in tension. The high sSFR regime will become increasingly important as future surveys push observations to higher redshifts. See, e.g., figure 4 of \cite{Behroozi2013} showing that galaxies at high redshift are predominantly high sSFR galaxies with an sSFR in excess of $\log{\mathrm{sSFR}} \gtrsim -10$. A significant part of these galaxies will have sSFR values above $\log{\mathrm{sSFR}} > -9$. It is difficult to determine the validity of the models of this work in this regime as there are only few constraints from observations at $\log{\mathrm{sSFR}} > -9$. The reliability of the predicted sSNR in this regime will therefore be strongly dependent on the validity of the underlying assumed model, in this case specifically the DTD assumed in equation \ref{eq:dtd}. Conversely, observations at high sSFR will constrain the DTD in this regime.\\

The test results quoted in table \ref{tab:modelfits} corroborate the points made above, with an important additional point being that although the quoted uncertainties for the parameters of the models developed in this work are only estimates, these uncertainty estimates have no influence on the test results as the tests solely depend on the goodness-of-fit of the maximum likelihood parameters. Based on the above results and discussion we recommend that either the piecewise model or the smooth logarithm parametrisation developed in this work is used when predicting type Ia supernova rates.

\section{Acknowledgements}
The authors would like to thank Teddy Frederiksen for input and Patrick Kelly, Martin Sparre, and Tamara Davis for useful input and discussions. PA would like to thank Martin Sparre, Mathew Smith, Mark Sullivan, and Yan Gao for providing access to data. JH was supported by a VILLUM FONDEN Investigator grant (project number 16599). 

\bibliographystyle{mnras}
\bibliography{SNRAB}
\appendix
\section{Additional Fitting Details}
\label{app:fittingdetails}

As discussed in the main body of this work we apply a grid method to derive maximum likelihood parameters for the models introduced in this work. Grid methods are typically computationally expensive, with the computational workload quickly increasing with the number of dimensions.  In our case, the maximum number of free parameters to fit in one model is four, which is the regime where MCMC methods such as {\sc Emcee} can be substantially cheaper computationally than grid methods. The MCMC approach was however found to be unsuitable, due to many local extrema effecting the marginalised likelihoods. The one dimensional marginalised likelihoods could for one parameter have a maximum at the global maximum likelihood while for others have a maximum at some value not near the global maximum likelihood region. This also meant that while the likelihood grid was utilised to find the maximum likelihood fit, it suffers from the same issues as the MCMC approach when used to estimate uncertainties. Therefore a bootstrapping approach was attempted. In this approach the the data is resampled with replacement and then refit multiple times, to derive distributions for all parameters from which uncertainties can be estimated. This proved to be computationally too expensive, as for each resampling of the data the grid method had to be rerun. Finally the Fisher matrix method was used to estimate uncertainties. First the Fisher matrix method was applied to the A+B model, which yielded uncertainties in agreement with those of the MCMC method to within 1\%. The Fisher matrix method was then used to estimate uncertainties for the piecewise model and the smooth logarithm parametrisation, respectively.\\

The results of the fit and code are available at https://github.com/per-andersen/SNR-AB.

\end{document}